\documentclass[twoside]{article}%
\usepackage{amssymb}
\usepackage{amsfonts}
\usepackage{amsmath}
\usepackage{graphicx}%
\providecommand{\U}[1]{\protect\rule{.1in}{.1in}}
\newtheorem{theorem}{Theorem}

\newtheorem{corollary}[theorem]{Corollary}

\newtheorem{lemma}[theorem]{Lemma}

\newtheorem{remark}[theorem]{Remark}

\newenvironment{proof}[1][Proof]{\noindent\textbf{#1.} }{\ \rule{0.5em}{0.5em}}
\begin{document}

\title{\textbf{Analytical Collapsing Solutions to Pressureless Navier-Stokes-Poisson
Equations with Density-dependent Viscosity }$\theta=1/2$ \textbf{in }$R^{2}$}
\author{Y\textsc{uen} M\textsc{anwai\thanks{E-mail address: nevetsyuen@hotmail.com }}\\\textit{Department of Applied Mathematics, }\\\textit{The Hong Kong Polytechnic University,}\\\textit{Hung Hom, Kowloon, Hong Kong}}
\date{Revised 24-Nov-2008v2}
\maketitle

\begin{abstract}
We study the 2-dimensional Navier--Stokes-Poisson equations with
density-dependent viscosity $\theta=1/2$ without pressure of gaseous stars in
astrophysics. The analytical solutions with collapsing in radial symmetry, are
constructed in this paper.

\end{abstract}

\section{Introduction}

The evolution of a self-gravitating fluid can be formulated by the
Navier-Stokes-Poisson equations of the following form:
\begin{equation}
\left\{
\begin{array}
[c]{rl}%
{\normalsize \rho}_{t}{\normalsize +\nabla\bullet(\rho u)} & {\normalsize =}%
{\normalsize 0,}\\
{\normalsize (\rho u)}_{t}{\normalsize +\nabla\bullet(\rho u\otimes u)+\nabla
P} & {\normalsize =}{\normalsize -\rho\nabla\Phi+vis(\rho,u),}\\
{\normalsize \Delta\Phi(t,x)} & {\normalsize =\alpha(N)}{\normalsize \rho,}%
\end{array}
\right.  \label{Euler-Poisson}%
\end{equation}
where $\alpha(N)$ is a constant related to the unit ball in $R^{N}$:
$\alpha(1)=2$; $\alpha(2)=2\pi$ and For $N\geq3,$%
\begin{equation}
\alpha(N)=N(N-2)V(N)=N(N-2)\frac{\pi^{N/2}}{\Gamma(N/2+1)},
\end{equation}
where $V(N)$ is the volume of the unit ball in $R^{N}$ and $\Gamma$ is a Gamma
function. And as usual, $\rho=\rho(t,x)$ and $u=u(t,x)\in\mathbf{R}^{N}$ are
the density, the velocity respectively. $P=P(\rho)$\ is the pressure.

In the above system, the self-gravitational potential field $\Phi=\Phi
(t,x)$\ is determined by the density $\rho$ through the Poisson equation.

And $vis(\rho,u)$ is the viscosity function:%
\begin{equation}
vis(\rho,u)=\bigtriangledown(\mu(\rho)\bigtriangledown\bullet u).
\end{equation}
Here we under a common assumption for:
\begin{equation}
\mu(\rho)\doteq\kappa\rho^{\theta}%
\end{equation}
and $\kappa$ and $\theta\geq0$ are the constants. In particular, when
$\theta=0$, it returns the expression for the $u$ dependent only viscosity
function:%
\begin{equation}
vis(\rho,u)=\kappa\Delta u.
\end{equation}
And the vector Laplacian in $u(t,r)$ can be expressed:%
\begin{equation}
\Delta u=u_{rr}+\frac{N-1}{r}u_{r}-\frac{N-1}{r^{2}}u.
\end{equation}
The equations (\ref{Euler-Poisson})$_{1}$ and (\ref{Euler-Poisson})$_{2}$
$(vis(\rho,u)\neq0)$ are the compressible Navier-Stokes equations with forcing
term. The equation (\ref{Euler-Poisson})$_{3}$ is the Poisson equation through
which the gravitational potential is determined by the density distribution of
the density itself. Thus, we call the system (\ref{Euler-Poisson}) the
Navier-Stokes-Poisson equations.

Here, if the $vis(\rho,u)=0$, the system is called the Euler-Poisson
equations.\ In this case, the equations can be viewed as a prefect gas model.
For $N=3$, (\ref{Euler-Poisson}) is a classical (nonrelativistic) description
of a galaxy, in astrophysics. See \cite{C}, \cite{KW} for a detail about the system.

$P=P(\rho)$\ is the pressure. The $\gamma$-law can be applied on the pressure
$P(\rho)$, i.e.%
\begin{equation}
{\normalsize P}\left(  \rho\right)  {\normalsize =K\rho}^{\gamma}\doteq
\frac{{\normalsize \rho}^{\gamma}}{\gamma}, \label{gamma}%
\end{equation}
which is a commonly the hypothesis. The constant $\gamma=c_{P}/c_{v}\geq1$,
where $c_{P}$, $c_{v}$\ are the specific heats per unit mass under constant
pressure and constant volume respectively, is the ratio of the specific heats,
that is, the adiabatic exponent in (\ref{gamma}). In particular, the fluid is
called isothermal if $\gamma=1$. With $K=0$, we call the system is pressureless.

For the $3$-dimensional case, we are interested in the hydrostatic equilibrium
specified by $u=0$. According to \cite{C}, the ratio between the core density
$\rho(0)$ and the mean density $\overset{\_}{\rho}$ for $6/5<\gamma<2$\ is
given by%
\begin{equation}
\frac{\overset{\_}{\rho}}{\rho(0)}=\left(  \frac{-3}{z}\dot{y}\left(
z\right)  \right)  _{z=z_{0}}%
\end{equation}
where $y$\ is the solution of the Lane-Emden equation with $n=1/(\gamma-1)$,%
\begin{equation}
\ddot{y}(z)+\dfrac{2}{z}\dot{y}(z)+y(z)^{n}=0,\text{ }y(0)=\alpha>0,\text{
}\dot{y}(0)=0,\text{ }n=\frac{1}{\gamma-1},
\end{equation}
and $z_{0}$\ is the first zero of $y(z_{0})=0$. We can solve the Lane-Emden
equation analytically for%
\begin{equation}
y_{anal}(z)\doteq\left\{
\begin{array}
[c]{ll}%
1-\frac{1}{6}z^{2}, & n=0;\\
\dfrac{\sin z}{z}, & n=1;\\
\dfrac{1}{\sqrt{1+z^{2}/3}}, & n=5,
\end{array}
\right.
\end{equation}
and for the other values, only numerical values can be obtained. It can be
shown that for $n<5$, the radius of polytropic models is finite; for $n\geq5$,
the radius is infinite.

Gambin \cite{G} and Bezard \cite{B} obtained the existence results about the
explicitly stationary solution $\left(  u=0\right)  $ for $\gamma=6/5$ in
Euler-Poisson equations$:$%
\begin{equation}
\rho=\left(  \frac{3KA^{2}}{2\pi}\right)  ^{5/4}\left(  1+A^{2}r^{2}\right)
^{-5/2}, \label{stationsoluionr=6/5}%
\end{equation}
where $A$ is constant.\newline The Poisson equation (\ref{Euler-Poisson}%
)$_{3}$ can be solved as%
\begin{equation}
{\normalsize \Phi(t,x)=}\int_{R^{N}}G(x-y)\rho(t,y){\normalsize dy,}%
\end{equation}
where $G$ is the Green's function for the Poisson equation in the
$N$-dimensional spaces defined by
\begin{equation}
G(x)\doteq\left\{
\begin{array}
[c]{ll}%
|x|, & N=1;\\
\log|x|, & N=2;\\
\dfrac{-1}{|x|^{N-2}}, & N\geq3.
\end{array}
\right.
\end{equation}
In the following, we always seek solutions in radial symmetry. Thus, the
Poisson equation (\ref{Euler-Poisson})$_{3}$ is transformed to%
\begin{equation}
{\normalsize r^{N-1}\Phi}_{rr}\left(  {\normalsize t,x}\right)  +\left(
N-1\right)  r^{N-2}\Phi_{r}{\normalsize =}\alpha\left(  N\right)
{\normalsize \rho r^{N-1},}%
\end{equation}%
\[
\Phi_{r}=\frac{\alpha\left(  N\right)  }{r^{N-1}}\int_{0}^{r}\rho
(t,s)s^{N-1}ds
\]

In this paper, we concern the analytical solutions with core collapsing for
the $2$-dimensional pressureless Navier-Stokes-Poisson equations with the
density-dependent viscosity. And our aim is to construct a family of such core
collapsing solutions.

Historically in astrophysics, Goldreich and Weber \cite{GW} constructed the
analytical collapsing (blowup) solution of the $3$-dimensional Euler-Poisson
equations for $\gamma=4/3$ for the non-rotating gas spheres. After that,
Makino \cite{M1} obtained the rigorously mathematical proof of the existence
of such kind of collapsing solutions. And in \cite{DXY}, the extension of the
above collapsing solutions to the higher dimensional cases ($N\geq4$). After
that, for the $2$-dimensional case, Yuen constructed the analytical collapsing
solutions for $\gamma=1$, in \cite{Y1}.

For the construction of the analytical solutions to the Navier-Stokes
equations in $R^{N}$, the Navier-Stokes-Poisson equations in $R^{3}$ without
pressure with $\theta=1$ and in $R^{4}$ without pressure with $\theta=5/4$,
readers may refer Yuen's recent results in \cite{Y2}, \cite{Y3}, \cite{Y4} respectively.

In this article, the analytical collapsing solutions are constructed in the
pressureless Navier--Stokes-Poisson equations with density-dependent viscosity
in $R^{2}$, with $\theta=1/2$, in radial symmetry:%
\begin{equation}
\left\{
\begin{array}
[c]{rl}%
\rho_{t}+u\rho_{r}+\rho u_{r}+{\normalsize \dfrac{1}{r}\rho u} &
{\normalsize =0,}\\
\rho\left(  u_{t}+uu_{r}\right)   & {\normalsize =-}\dfrac{2\pi\rho}{r}%
{\displaystyle\int_{0}^{r}}
\rho(t,s)sds+[\kappa\rho^{1/2}]_{r}u_{r}+(\kappa\rho^{1/2})(u_{rr}+\dfrac
{1}{r}u_{r}-\dfrac{1}{r^{2}}u),
\end{array}
\right.  \label{gamma=1}%
\end{equation}
in the form of the following theorem.

\begin{theorem}
\label{thm:1}For the $2$-dimensional pressureless Navier--Stokes-Poisson
equations with $\theta=1/2$, in radial symmetry, (\ref{gamma=1}), there exists
a family of solutions,%
\begin{equation}
\left\{
\begin{array}
[c]{c}%
\rho(t,r)=\dfrac{1}{(T-Ct)^{2}y(\frac{r}{T-Ct})^{2}},\text{ }%
{\normalsize u(t,r)=}\dfrac{-C}{T-Ct}{\normalsize r;}\\
\ddot{y}(z){\normalsize +}\dfrac{1}{z}\dot{y}(z)-{\normalsize \dfrac{2\pi
\sqrt{\kappa}}{C}\frac{1}{y(z)^{2}}=0,}\text{ }y(0)=\alpha>0,\text{ }\dot
{y}(0)=0,
\end{array}
\right.  \label{solution1}%
\end{equation}
where $T>0$, $\kappa>0$, $C\neq0$ and $\alpha$ are constants.\newline In
particular, for $C>0$, the solutions collapse in the finite time $T/C$.
\end{theorem}

\section{Separable Blowup Solutions}

Before presenting the proof of Theorem \ref{thm:1}, we prepare some lemmas.
First, we obtain the solutions for the continuity equation of mass in radial
symmetry (\ref{gamma=1})$_{1}$.

\begin{lemma}
\label{lem:generalsolutionformasseq}%
\label{lem:generalsolutionformasseq copy(1)}For the equation of conservation
of mass in radial symmetry:
\begin{equation}
\rho_{t}+u\rho_{r}+\rho u_{r}+\frac{1}{r}\rho u=0,\label{eq1212}%
\end{equation}
there exist solutions,%
\begin{equation}
\rho(t,r)=\frac{1}{\left(  T-Ct\right)  ^{2}y(\frac{r}{T-Ct})^{2}},\text{
}{\normalsize u(t,r)=}\frac{-C}{T-Ct}{\normalsize r,}\label{eq1313}%
\end{equation}
with the form $y\neq0$ and $y\in C^{1}$, $C$ and $T>0$ are constants$.$
\end{lemma}

\begin{proof}
We just plug (\ref{eq1313}) into (\ref{eq1212}). Then
\begin{align*}
&  \rho_{t}+u\rho_{r}+\rho u_{r}+\frac{1}{r}\rho u\\
&  =\frac{(-2)(-C)}{\left(  T-Ct\right)  ^{3}y(\frac{r}{T-Ct})^{2}}%
+\frac{(-2)(-1)(-C)r\overset{\cdot}{y}(\frac{r}{T-Ct})}{\left(  T-Ct\right)
^{4}y(\frac{r}{T-Ct})^{3}}\\
&  +\frac{(-C)r}{T-Ct}\frac{(-2)}{(T-Ct)^{2}y(\frac{r}{T-Ct})^{3}}%
\frac{\overset{\cdot}{y}(\frac{r}{T-Ct})}{T-Ct}+\frac{1}{(T-Ct)^{2}y(\frac
{r}{T-Ct})^{2}}\frac{(-C)}{T-Ct}\\
&  +\frac{1}{r}\frac{1}{(T-Ct)^{2}y(\frac{r}{T-Ct})^{2}}\frac{(-C)}{T-Ct}r\\
&  =0.
\end{align*}
The proof is completed.
\end{proof}

Besides, we need the lemma for stating the property of the function $y(z)$.
The similar lemma was already given in Lemmas 9 and 10, \cite{Y1}, by the
fixed point theorem. For the completeness, the proof is also presented here.

\begin{lemma}
\label{lemma2}For the ordinary differential equation,%
\begin{equation}
\left\{
\begin{array}
[c]{c}%
\ddot{y}(z){\normalsize +}\dfrac{1}{z}\dot{y}(z)-{\normalsize \frac{\sigma
}{y(z)^{2}}=0}\\
y(0)=\alpha>0,\text{ }\dot{y}(0)=0,
\end{array}
\right.  \label{SecondorderElliptic}%
\end{equation}
where $\sigma$ is a positive constant,\newline has a solution $y(z)\in C^{2}$
and $\underset{z\rightarrow+\infty}{\lim}y(z)=\infty$.
\end{lemma}

\begin{proof}
By integrating (\ref{SecondorderElliptic}), we have,%
\begin{equation}
\overset{\cdot}{y}(z)=\frac{\sigma}{z}\int_{0}^{z}\frac{1}{y(s)^{2}}%
sds\geq0.\label{lemma3eq1}%
\end{equation}
Thus, for $0<z<z_{0}$, $y(x)$ has a uniform lower upper bound
\[
y(z)\geq y(0)=\alpha>0.
\]
As we obtained the local existence in Lemma \ref{lemma2}, there are two
possibilities:\newline(1)$y(z)$ only exists in some finite interval
$[0,z_{0}]$: (1a)$\underset{z\rightarrow z_{0-}}{\lim}y(z)=\infty$; (1b)$y(z)$
has an uniformly upper bound, i.e. $y(z)\leq\alpha_{0}$ for some constant
$\alpha_{0}.$\newline(2)$y(z)$ exists in $[0,$ $+\infty)$: (2a)$\underset
{z\rightarrow+\infty}{\lim}y(z)=\infty$; (2b)$y(z)$ has an uniformly upper
bound, i.e. $y(z)\leq\beta$ for some positive constant $\beta$.\newline We
claim that possibility (1) does not exist. We need to reject (1b) first: If
the statement (1b) is true, (\ref{lemma3eq1}) becomes%
\begin{equation}
\frac{\sigma z}{2\alpha^{2}}=\frac{\sigma}{z}\int_{0}^{z}\frac{s}{\alpha^{2}%
}ds\geq\overset{\cdot}{y}(z).\label{possible1}%
\end{equation}
Thus, $\overset{\cdot}{y}(z)$ is bounded in $[0,z_{0}]$. Therefore, we can use
the fixed point theorem again to obtain a large domain of existence, such that
$[0,z_{0}+\delta]$ for some positive number $\delta$. There is a
contradiction. Therefore, (1b) is rejected.\newline Next, we do not accept
(1a) because of the following reason: It is impossible that $\underset
{z\rightarrow z_{0-}}{\lim}y(z)=\infty$, as from (\ref{possible1}),
$\overset{\cdot}{y}(z)$ has an upper bound in $[0,$ $z_{0}]$:%
\begin{equation}
\frac{\sigma z_{0}}{2\alpha^{2}}\geq\overset{\cdot}{y}(z).\label{lemma3eq2}%
\end{equation}
Thus, (\ref{lemma3eq2}) becomes,
\begin{align*}
y(z_{0}) &  =y(0)+\int_{0}^{z_{0}}\overset{\cdot}{y}(s)ds\\
&  \leq\alpha+\int_{0}^{z_{0}}\frac{\sigma z_{0}}{2\alpha^{2}}ds\\
&  =\alpha+\frac{\sigma(z_{0})^{2}}{2\alpha^{2}}%
\end{align*}
Since $y(z)$ is bounded above in $[0,$ $z_{0}]$, it contracts the statement
(1a), such that $\underset{z\rightarrow z_{0-}}{\lim}y(z)=\infty$. So, we can
exclude the possibility (1).\newline We claim that the possibility (2b)
doesn't exist. It is because
\[
\overset{\cdot}{y}(z)=\frac{\sigma}{z}\int_{0}^{z}\frac{s}{y(s)^{2}}%
ds\geq\frac{\sigma}{z}\int_{0}^{z}\frac{s}{\beta^{2}}ds=\frac{\sigma z}%
{2\beta^{2}}.
\]
Then, we have,%
\begin{equation}
y(z)\geq\alpha+\frac{\sigma}{4\beta^{2}}z^{2}.\label{lemma3eq3}%
\end{equation}
By letting $z\rightarrow\infty$, (\ref{lemma3eq3}) turns out to be,
\[
y(z)=\infty.
\]
Since a contradiction is established, we exclude the possibility (2b). Thus,
the equation (\ref{SecondorderElliptic}) exists in $[0,$ $+\infty)$ and
$\underset{z\rightarrow+\infty}{\lim}y(z)=\infty$. This completes the proof.
\end{proof}

Here we are already to give the proof of Theorem \ref{thm:1}.

\begin{proof}
[Proof of Theorem 2]From Lemma \ref{lem:generalsolutionformasseq}, it is clear
for that (\ref{solution1}) satisfy (\ref{gamma=1})$_{1}$. For the momentum
equation (\ref{gamma=1})$_{2}$, we get,%
\begin{align}
&  \rho(u_{t}+uu_{r})+\frac{2\pi\rho}{r}%
{\displaystyle\int\limits_{0}^{r}}
\rho(t,s)sds-[\mu(\rho)]_{r}u_{r}-\mu(\rho)(u_{rr}+\dfrac{1}{r}u_{r}-\dfrac
{1}{r^{2}}u)\\
&  =\rho\left[  \frac{(-C)(-1)(-C)}{(T-Ct)^{2}}r+\frac{(-C)}{T-Ct}r\cdot
\frac{(-C)}{T-Ct}\right]  +\frac{2\pi\rho}{r}%
{\displaystyle\int\limits_{0}^{r}}
\frac{1}{(T-Ct)^{2}y(\frac{r}{T-Ct})^{2}}sds\\
&  -(\kappa\rho^{1/2})_{r}\frac{(-C)}{T-Ct}-\mu(\rho)\left(  0+\dfrac{1}%
{r}\frac{(-1)}{T-Ct}-\frac{1}{r^{2}}\frac{(-1)}{T-Ct}r\right)  \nonumber\\
&  =\frac{\partial}{\partial r}\left[  \frac{\kappa}{(T-Ct)^{2}y(\frac
{r}{C-ct})^{2}}\right]  ^{1/2}\frac{C}{T-Ct}+\frac{2\pi\rho}{r}%
{\displaystyle\int\limits_{0}^{r}}
\frac{1}{(T-Ct)^{2}y(\frac{r}{T-Ct})^{2}}sds\\
&  =\frac{1}{2}\left[  \frac{\kappa}{(T-Ct)^{2}y(\frac{r}{T-Ct})^{2}}\right]
^{-1/2}\frac{(-2)}{(T-Ct)^{2}y(\frac{r}{T-Ct})^{3}}\frac{\dot{y}(\frac
{r}{T-Ct})}{T-Ct}\frac{C}{T-Ct}\nonumber\\
&  +\frac{2\pi\rho}{r}%
{\displaystyle\int\limits_{0}^{r}}
\frac{1}{(T-Ct)^{2}y(\frac{s}{T-Ct})^{2}}sds\\
&  =\frac{-C}{\sqrt{\kappa}}\frac{\rho\dot{y}(\frac{r}{T-Ct})}{(T-Ct)}%
+\frac{2\pi\rho}{r}%
{\displaystyle\int\limits_{0}^{r}}
\frac{1}{(T-Ct)^{2}y(\frac{s}{T-Ct})^{2}}sds\\
&  =\frac{-\rho}{(T-Ct)}\left[  \frac{C}{\sqrt{\kappa}}\dot{y}(\frac{r}%
{T-Ct})-\frac{2\pi}{r(T-Ct)}%
{\displaystyle\int\limits_{0}^{r}}
\frac{1}{y(\frac{s}{T-Ct})^{2}}sds\right]  \\
&  =\frac{-\rho}{(T-Ct)}\left[  \frac{C}{\sqrt{\kappa}}\dot{y}(\frac{r}%
{T-Ct})-\frac{2\pi}{\frac{r}{T-Ct}}%
{\displaystyle\int\limits_{0}^{r/(T-Ct)}}
\frac{1}{y(\tau)^{2}}\tau d\tau\right]  .
\end{align}
By letting $\tau=r/(T-Ct)$, it follows:%
\begin{align}
&  =\frac{-\rho}{(T-Ct)}\left[  \frac{C}{\sqrt{\kappa}}\dot{y}(\frac{r}%
{T-Ct})-\frac{2\pi}{\frac{r}{T-Ct}}%
{\displaystyle\int\limits_{0}^{r/(T-Ct)}}
\frac{1}{y(\tau)^{2}}\tau d\tau\right]  \\
&  =\frac{-\rho}{(T-Ct)}Q\left(  \frac{r}{T-Ct}\right)  .\nonumber
\end{align}
And denote $z=r/(T-ct)$,%
\[
Q(\frac{r}{T-Ct})={\normalsize Q(z)=}\frac{C}{\sqrt{\kappa}}\dot
{y}(z){\normalsize -}\dfrac{2\pi}{z}%
{\displaystyle\int\limits_{0}^{z}}
\frac{1}{y(\tau)^{2}}\tau d\tau{\normalsize .}%
\]
Differentiate $Q(z)$\ with respect to $z$,%
\begin{align*}
\dot{Q}(z) &  =\frac{C}{\sqrt{\kappa}}\ddot{y}(z)-\frac{{\normalsize 2\pi}%
}{y(z)^{2}}+\frac{2\pi}{z^{2}}%
{\displaystyle\int\limits_{0}^{z}}
\frac{1}{y(\tau)^{2}}\tau{\normalsize d\tau}\\
&  =-\frac{C}{\sqrt{\kappa}}\frac{\dot{y}(z)}{z}+\frac{2\pi}{z^{2}}%
{\displaystyle\int\limits_{0}^{z}}
\frac{1}{y(\tau)^{2}}\tau{\normalsize d\tau}\\
&  =-\frac{1}{z}Q(z),
\end{align*}
where the above result is due to the fact that we choose the following
ordinary differential equation:%
\[
\left\{
\begin{array}
[c]{c}%
\ddot{y}(z){\normalsize +}\dfrac{1}{z}\dot{y}(z)-{\normalsize \dfrac{2\pi
\sqrt{\kappa}}{C}\frac{1}{y(z)^{2}}=0}\\
{\normalsize y(0)=\alpha>0,}\text{ }\dot{y}(0){\normalsize =0.}%
\end{array}
\right.
\]
With $Q(0)=0$, this implies that $Q(z)=0$. Thus, the momentum equation
(\ref{gamma=1})$_{2}$ is satisfied.\newline With Lemma \ref{lemma2} about
$y(z)$, we are able to show that the family of the solutions collapse in
finite time $T/C$. This completes the proof.
\end{proof}

The statement about the blowup rate will be immediately followed:

\begin{corollary}
The collapsing rate of the solution (\ref{solution1}) is
\begin{equation}
\underset{t\rightarrow T/C^{-}}{\lim}\rho(t,0)(T-Ct)^{2}\geq O(1).
\end{equation}

\end{corollary}

\begin{remark}
Besides, if we consider the $2$-dimensional Navier-Stokes equations with the
repulsive force in radial symmetry with $\theta=1/2$,%
\begin{equation}
\left\{
\begin{array}
[c]{rl}%
\rho_{t}+u\rho_{r}+\rho u_{r}+{\normalsize \dfrac{1}{r}\rho u} &
{\normalsize =0,}\\
\rho\left(  u_{t}+uu_{r}\right)  & {\normalsize =+}\dfrac{2\pi\rho}{r}\int
_{0}^{r}\rho(t,s)sds+[\kappa\rho^{1/2}]u_{r}+(\kappa\rho^{1/2})(u_{rr}%
+\dfrac{1}{r}u_{r}-\dfrac{1}{r^{2}}u),
\end{array}
\right.
\end{equation}
the special solutions are:%
\begin{equation}
\left\{
\begin{array}
[c]{c}%
\rho(t,r)=\dfrac{1}{(T-Ct)^{2}y\left(  \frac{r}{T-Ct}\right)  ^{2}}\text{,
}{\normalsize u(t,r)=}\dfrac{-C}{T-Ct}{\normalsize r}\\
\ddot{y}(z){\normalsize +}\dfrac{1}{z}\dot{y}(z)+{\normalsize \dfrac{2\pi
\sqrt{\kappa}}{C}\frac{1}{y(z)^{2}}=0},\text{ }y(0)=\alpha\neq0,\text{ }%
\dot{y}(0)=0,
\end{array}
\right.
\end{equation}

\end{remark}


\begin{thebibliography}{99}                                                                                               %


\bibitem {B}M. Bezard, \textit{Existence locale de solutions pour les
equations d'Euler-Poisson. (French) [Local Existence of Solutions for
Euler-Poisson Equations]} Japan J. Indust. Appl. Math. \textbf{10} (1993), no.
3, 431--450.

\bibitem {C}S. Chandrasekhar, \textit{An Introduction to the Study of Stellar
Structure}, Univ. of Chicago Press, 1939.

\bibitem {DXY}Y.B. Deng, J.L. Xiang and T. Yang, \textit{Blowup Phenomena of
Solutions to Euler-Poisson Equations}, J. Math. Anal. Appl. \textbf{286}
(1)(2003), 295-306.

\bibitem {G}P. Gamblin, \textit{Solution reguliere a temps petit pour
l'equation d'Euler-Poisson. (French) [Small-time Regular Solution for the
Euler-Poisson Equation]} Comm. Partial Differential Equations \textbf{18}
(1993), no. 5-6, 731--745.

\bibitem {GW}P.Goldreich, S. Weber, \textit{Homologously Collapsing Stellar
Cores}, Astrophys, J. \textbf{238}, 991 (1980).

\bibitem {KW}R. Kippenhahn, A,Weigert, \textit{Stellar Sturture and
Evolution}, Springer-Verlag, 1990.

\bibitem {M1}T. Makino, \textit{Blowing up Solutions of the Euler-Poission
Equation for the Evolution of the Gaseous Stars}, Transport Theory and
Statistical Physics \textbf{21} (1992), 615-624.

\bibitem {Y1}M.W. Yuen, \textit{Analytical Blowup Solutions to the
2-dimensional Isothermal Euler-Poisson Equations of Gaseous Stars}, J. Math.
Anal. Appl. \textbf{341 (}1\textbf{)(}2008\textbf{), }445-456.

\bibitem {Y2}M.W. Yuen, \textit{Analyitcal Solutions to the Navier-Stokes
Equations}, Journal of Mathematical Physics, \textbf{49} (2008) No. 11,
113102, 10pp.

\bibitem {Y3}M. W. Yuen, \textit{Analytical Solutions to the 3-dimensional
Pressuless Navier-Stokes-Poisson Equations with Density-dependent Viscosity},
Submitted,\textbf{ }arXiv:0811.0379v1.

\bibitem {Y4}M. W. Yuen, \textit{Analytical Solutions to the 4-dimensional
Pressuless Navier-Stokes-Poisson Equations with Density-dependent Viscosity},
Submitted,\textbf{ }arXiv:0811.1323v1.
\end{thebibliography}
\end{document}